\documentstyle[12pt]{article}
\newcommand\beq{\begin{equation}}
\newcommand\eeq{\end{equation}}

\title{Colour Screening, Quark Propagation\\
in Nuclear Matter and\\
the Broadening of the Momentum Distribution\\
of Drell-Yan Pairs}
\author{J. Dolej\v s\' \i$^{1,2}$, J. H\"ufner$^1$, B.Z.Kopeliovich$^{1,3}$\\
\\
$^1$ Inst. of Theoretical Physics, University of Heidelberg\\
Philosophenweg 19, D-6900 Heidelberg\\
\\
$^2$Dept. of Nuclear Physics\\
Fac. of Mathematics and Physics, Charles University\\
V Hole\v sovi\v ck\'ach 2, CS-18000 Praha 8\\
\\
$^3$ Laboratory of Nuclear Problems\\
Joint Institute for Nuclear Research\\
Head Post Office, P.O. Box 79\\
101000 Moscow, Russia}

\date{ }

\begin{document}
\maketitle
\begin{abstract}
We calculate the broadening of the transverse momentum distribution of
a quark propagating through nuclear matter. Colour screening plays
a fundamental role in that it cuts off quark-nucleon interactions with
soft gluons. The mean transverse momentum of the quark acquired along its
trajectory, observed via Drell-Yan pairs, is related to
it the ratio of the total inelastic meson-nucleon cross section
it to the meson mean squared radius.
Parameter-free calculations agree with the data.
\end{abstract}

\newpage

Nuclei may be used as micro-laboratories for the observation of processes
which cannot be studied otherwise, e.g. strong interaction
dynamics at short-time intervals before or after an interaction.
This information is completely lost in the scattering on a nucleon. Nuclei
provide a unique opportunity to study properties of hadronic systems at early
stages of their formation. Propagation of a quark (antiquark) through
nuclear matter is one of the examples. The momentum distribution of such
a quark after the multiple interaction with the target nucleons
can be measured in a Drell-Yan process of lepton-antilepton
pair production. Since the lepton pair does not interact on its way out
of the nucleus, it carries the undistorted information about the
interactions of the quark.

In this paper, we study the observable of the transverse momentum,
particularly the broadening of the transverse momentum distribution of
Drell-Yan pairs produced on nuclei, as compared to a nucleon target.
Empirically, the increase of the mean squared transverse momentum
$<\!p^2_T\!>$ is proportional to $L_A$, the mean length of the quark path in a
nucleus before creation of the Drell-Yan pair.
\beq
\delta<\!p^2_T\!>_{DY}^A= \;<\!p^2_T\!>_{DY}^A-<\!p^2_T\!>_{DY}^p=\kappa\rho_A
L
\label{aa}\eeq
with $\kappa$ being a dimensionless and $A$-independent constant;
$\rho_A$ is the nuclear density. Relation (\ref{aa}) resembles broadening
originating from a ``random walk''-process.  It is the
purpose of our letter to derive eq.~(\ref{aa}) and to calculate the
constant $\kappa$, starting from basic available information about
strong interaction dynamics.

Confinement is a fundamental aspect of strong interactions. The colour of a
quark is always compensated by the colours of other accompanying partons.
While this fact is unimportant at high momentum transfers, it is crucial for
soft processes. Here the colour screening cuts off all low-momentum gluons
whose wavelength is longer than the screening-length scale. Mathematically,
this is equivalent to a cut-off of undesirable infrared divergences.

We aim at calculating the transverse momentum distribution of a
high-energy quark, which has traversed nuclear matter. We call this quark,
originating from a parent high-energy hadron, the tagged quark in distinction
to the screening partons which we call for simplicity ``antiquarks''.
At sufficiently high energies, the quark-antiquark separation is ``frozen''
by the Lorentz transformation, while the two partons pass the nucleus.
These physical ideas can be technically realized as follows.
The individual amplitudes of quark-nucleon scattering convolute to
a hadron-nucleus multiple-interaction amplitude, with the tagged quark
having in the final state an additional transverse
momentum $\vec q$, relative to the initial hadron momentum.
The corresponding diagram is shown in fig.~1.
Let us start with the $n$-fold scattering amplitude. Its square summed over
all final states of nucleons participating in the interaction and
integrated over all the transferred momenta, fixing the tagged-quark final
momentum $\vec q$, has the form:
\beq
\left(\frac{d\sigma}{d^2q}\right)_n=(2\pi)^2\int^1_0 \!\!d\alpha\prod^n_{i=1}
\frac{d^2k_i}{(2\pi)^2} \;f(k_i)\left|\sum^n_{m=0}
{n\choose m}(-1)^{n-m}\phi\biggl(\vec
q- \sum^m_{\ell=1}\vec k_\ell|\alpha\biggr)\right|^2.
\label{ab}\eeq
While $n$ gluons are exchanged between the target nucleons and the
incoming $q\overline{q}$ system, only $m \;(\leq n)$ gluons couple
to the tagged quark (c.f. fig.~1).
The wave function $\phi(\vec k|\alpha)$ of the incoming hadron depends
on the transverse momentum $\vec k$ of the tagged quark, and the relative
longitudinal momentum $0<\alpha<1$. The function $f(k)$ in eq. (\ref{ab})
includes the two-gluon-nucleon coupling and gluon propagators
\beq
f(k)=\frac{8}{3}\;\frac{\alpha^2_s(k)}{(k^2+m^2_g)^2}\;
\bigl\lbrack1-F_N(k^2)\bigr\rbrack,\label{ac}
\eeq
where $F_N(k^2)$ is the two-quark formfactor of a nucleon:
$F_N(k^2)=<\!N|\exp(i\vec k(\vec r_1-\vec r_2))|N\!\!>$.
The effective gluon mass $m_g$ is introduced to take into account
confinement and is inessential at large~$k$. We use the one-loop
approximation for the QCD coupling constant $\alpha_s(k)$ at large~$k$, and
freeze it at small~$k$. The details can be found in \cite{kop}.

After transformating the wave functions $\phi(\vec k|\alpha)$ in
eq.~(\ref{ab}) to the
mixed $\vec\rho,\alpha$-representation, where $\vec\rho$ is a
quark-antiquark separation in the impact parameter plane, eq.~(\ref{ab})
changes into
\beq
\left(\frac{d\sigma}{d^2q}\right)_n=\frac{1}{(2\pi)^2}\int^1_0\!\! d\alpha
\int \!d^2\!\rho\:d^2\!\rho\,'\ e^{i\vec q\,(\vec\rho-\vec\rho\,')}
\phi(\vec\rho|\alpha)\:\phi^*(\vec\rho\,'|\alpha)
\left\lbrack\sigma(\vec\rho,\vec\rho\,')\right\rbrack^n.
\label{ad}\eeq
Here
\beq
\sigma(\vec\rho,\vec\rho\,')=\int\!\frac{d^2k}{(2\pi)^2}\:f(k)\left(1-
e^{-i\vec k\vec\rho}\right)\left(1-e^{i\vec
k\vec\rho\,'}\right).
\label{ae}\eeq
Note that $\sigma(\vec\rho,\vec\rho\,')$ can be represented as
\beq
\sigma(\vec\rho,\vec\rho\,')=\frac{1}{2} \sigma(\vec\rho) +
\frac{1}{2}\sigma(\vec\rho\,') - \frac{1}{2}\sigma(\vec\rho-\vec\rho\,'),
\label{af}\eeq
where
\beq
\sigma(\vec\rho)=2 \int\!\frac{d^2k}{(2\pi)^2}\:f(k^2)\left(1-e^{-i\vec
k\vec\rho}\right)
\label{ag}\eeq
is the inelastic interaction cross section of $\bar qq$-pair with separation
$\rho$, with a nucleon \cite{zam}.

We sum eq. (\ref{ad}) over $n$ with weight factors $T^n(b)/n!$,
where $T(b)$ is the nuclear thickness function at the impact parameter
$b,\ T(b) =\int^\infty_{-\infty}\rho_A(b,z)\,dz$.
In the Drell-Yan process one has to include the term with $n\!=\!0$,
which corresponds to no initial state interaction of the incoming hadron
in the nucleus before lepton-pair production. One also has to take into
account the absorptive corrections, which in $b$-representation have a
simple form,
$\exp\{-\frac{1}{2}\lbrack\sigma(\vec\rho)+\sigma(\vec\rho\,')\rbrack \,
T(b)\}$ being the probability amplitude to have no inelastic interactions
in the initial and final state.

Putting the factors together and using eq.~(\ref{af}) we get
\beq
\frac{d\sigma}{d^2q}=\frac{1}{(2\pi)^2}\int\! d^2b\int^1_0\!\! d\alpha
\int\! d^2\!\rho\: d^2\!\rho\,'e^{i\vec q\,(\vec\rho-\vec\rho\,')}
\phi(\vec\rho|\alpha)\:\phi^*(\vec\rho\,'|\alpha)\:e^{-\frac{1}{2}\sigma
(\vec\rho-\vec\rho\,')\:T(b)}.
\label{ah} \eeq
The integration over $\vec q$ at fixed impact parameter gives 1,
since $\sigma(0)\!=\!0$, eq.~(\ref{ag}). This result could have been
anticipated
since the quark does not disappear in the integration of all momenta.

Expression eq.~(\ref{ah}) is the main technical result of this paper.
It is of more general nature than the double-gluon approximation
used as an input. One can use any QCD-inspired model for $\sigma(\vec\rho)$.
Nevertheless we will use the results of the two-gluon approximation
with running coupling $\alpha_s(k)$, of ref.~\cite{kop}.
For small $\rho^2, \; \rho^2\ll 1$~fm$^2$, where perturbative QCD is valid,
one gets
\beq
\sigma(\vec\rho)=C\rho^2,
\label{ba} \eeq
with a dimensionless constant $C=3.2$ \cite{kop}.
The limit of small $\rho$ is given in Drell-Yan production of heavy lepton
pair to be discussed below, since the virtuality of the tagged quark is
of the order of the squared mass of the pair.

If in addition we use a Gaussian approximation for the hadron wave function
in transverse direction
$\phi(\vec r)=\sqrt{\mu^2_h/\pi}\exp(-\frac{1}{2}\mu^2_h
r^2)$, we obtain the cross section eq. (\ref{ah}) in the explicit form
\beq
\frac{d\sigma}{d^2q}=\int\! d^2b\:\frac{1}{\pi\lbrack\mu^2_h+2CT(b)\rbrack}
\:e^{-q^2/\lbrack\mu^2_h+2CT(b)\rbrack}.
\label{ai}\eeq

With the help of eq.~(\ref{ai}), one can derive an expression
for the cross section of the Drell-Yan process on a nucleus. We combine
the transverse momentum distribution of the tagged projectile quark after it
has
traversed the nuclear thickness $T(b,z)=\int_{-\infty}^z dz'\rho_A(b,z')$
with the transverse momentum distribution of an antiquark in a target
nucleon to obtain
\beq
\frac{d\sigma^{hA}_{DY}}{d^2 q}=\sigma^{hN}_{DY}\int\! d^2b
\int^\infty_{-\infty}\!\! dz\:\rho_A(b,z)\:
\frac{1}{\pi\lbrack<\!p^2_T\!>^{hN}_{DY}+2CT(b,z)\rbrack}
e^{-q^2/\lbrack<\!p^2_T\!>^{hN}_{DY}+2CT(b,z)\rbrack}.
\label{aj}\eeq
Here, $\sigma^{hN}_{DY}$ is the
total cross section of the Drell-Yan process on a nucleon and
$<\!p^2_T\!>^{hN}_{DY}$ the mean squared transverse momentum of the lepton
pair produced in a $hN$ collision.

One deduces from eq.~(\ref{aj}) that the increase of the mean squared
transverse momentum of the lepton pair produced on a nucleus compared to
a nucleon as target, is
\beq
\delta<\!p^2_T\!>^{hA}_{DY}=<\!p^2_T\!>^{hA}_{DY}-<\!p^2_T\!>^{hN}_{DY}=
C<\!T\!>,
\label{ak}\eeq
where $<\! T\!>=\frac{1}{A}\int\! d^2b\;T^2(b)$, is the mean nuclear
thickness. The result eq.~(\ref{ak}) confirms the empirical observation
mentioned earlier eq.~(\ref{aa}), that the broadening of
$<\!p^2_T\!>^{hA}_{DY}$ of a lepton-pair produced on a nucleus
is proportional to the mean length of path of the projectile quark
in the nucleus, $L_A=<\!T\!>/2\rho_A$. Moreover,
\beq
\kappa=2C.
\label{bb}\eeq
Eq.~(\ref{bb} relates the constant of proportionality $\kappa$ in
eq.~(\ref{aa}) for the mean squared momentum transfer $\delta\!<\!p^2_T\!>$
to the proportionality constant $C$, which governs the inelastic cross
section $\sigma(\rho)=C\rho^2$ eq.~(\ref{ba}) for the interaction
of a $q\overline q$ pair transverse separation $\rho$ with a nucleon.
It is remarkable that according to eq.~(\ref{bb}) the constant $\kappa$
is independent of properties of the colliding hadrons and of the kinematics
of the Drell-Yan process.

The experiment NA10 \cite{bor} has measured a value
\beq
\delta<\!p^2_T\!>^{\pi W}_{DY}=0.15\pm.03\mbox{stat}\pm.03\mbox{syst (GeV/c)}^2
\label{al}\eeq
for incoming pions at 140 and 286~GeV, while the experiment E772 with
800~GeV protons \cite{ald} reports a value
\beq
\delta<\!p^2_T\!>^{pW}_{DY}=0.113\pm.016\ \mbox{(GeV/c)}^2.
\label{am}\eeq
The two values coincide within the error bars which implies that
no dependence on the type of incident hadron and its energy is visible,
as expected from our expression eq.~(\ref{bb}). The expression
eq.~(\ref{ak}) predicts a value
\beq
\delta<\!p^2_T\!>^{hW}_{DY}=0.17 \mbox{(GeV/c)}^2
\label{an}\eeq
for $C=3.2$ from \cite{kop} in fair agreement with the data.

We calculate the ratio of differential cross sections of the Drell-Yan
process on nucleus to nucleon targets, $R(A/N)$, normalized with atomic
weight $A$. We use expression (\ref{aj}) with $<\!p^2_T\!>^{hN}_{DY}$
from the systematics of \cite{GPS} of D-Y data.
\begin{eqnarray}
<\!p^2_T\!>^{\pi p}_{DY}&=&0.59+0.0029\: s\nonumber\\
<p_T>^{pp}_{DY}&=&0.39+0.028 \,\sqrt s.
\label{ao} \end{eqnarray}
Here, $s$ is the square of the total c.m. energy in $GeV^2$.
Assuming a Gaussian form for the $p_T$-distribution, we use
$<\!p^2_T\!>={4\over \pi} <p_T>^2$.

The results of calculations of $R(A/N)$ are compared with experimental data
\cite{bor,al2} on fig.~2. One can see that our parameter-free calculations
provide a good description of experimental data. The flattening of
the $p_T$-dependence of $R(A/N)$ at high energies observed experimentally
is a result of the increase of $<\!p^2_T\!>^{hN}_{DY}$ as a function
of energy.

\bigskip

\subsubsection*{Conclusions}
\begin{itemize}
\item The phenomenon of broadening of the transverse momentum
distribution of a quark propagating through nuclear matter crucially depends
on the effect of colour screening in soft interactions. While the cross
section $\sigma(\rho)$ for the individual quark-nucleon scattering
decreases with decreasing screening radius~$\rho$, the mean transverse
momentum picked up in each collision is inversely proportional to $\rho$.
The product of $<\!p^2_T\!>.\sigma(\rho)$ in each collision is therefore
independent of $\rho$ and equals the constant $2C$. The smaller the
screening radius, the rarer the quark rescatters, but the larger is mean
transverse momentum in each rescattering.
\item We calculate the cross section $\sigma(\rho)$ in the one-gluon exchange
approximation for the hadron inelastic scattering amplitude. This approach
is justified because of the high virtuality of the quarks participating
in the Drell-Yan reaction.
\item Analogous consideration may be applied to the hadroproduction of
charmonia on nuclei which predominantly proceed via gluon fusion.
We expect an approximate doubling of the broadening effect due to
the additional factor $9/4$ in the cross section of the gluon-nucleon
interaction. Experimental data on $J/\Psi$ production on nuclei at 200 GeV
confirm the expected increase and give values for
$\kappa_{J/\Psi}/\kappa_{DY}=1.9 \pm 0.4$ \cite{bor} and $1.6 \pm 0.25$
\cite{bag}. However, for the $\Upsilon$ production at 800 GeV \cite{ald}
the experimental ratio $\kappa_{J/\Psi}/\kappa_{DY}=5.9 \pm 1.3$
is much larger than what we would expect and needs another explanation.
\item The process of inclusive production of hadrons with high transverse
momenta off nuclei (``Cronin effect'') has much to do with the colour
screening effect. Indeed the high transverse momentum is built up of many
soft scatterings in a way similar to the case treated in this paper for
the Drell-Yan process.
\item The broadening of the transverse momentum of a quark propagating
through nuclear matter, might be relevant also to the process of inclusive
production of leading hadrons in deep inelastic scattering on nuclei.
However, it occures only if the hadron is produced outside the
nucleus,
i.e. the length of the formation zone exceeds the nuclear radius. Otherwise,
the broadening becomes much weaker than in the Drell-Yan process, as the
colourless hadron should reinteract in the nucleus with a small elastic
cross section.
\end{itemize}

\bigskip
{\bf Acknowledgements:} J. D. and B. Z. K. gratefully acknowledge
the financial support of their visit in Heidelberg by the
Max-Planck-Institut f\"ur Kernphysik. The work has been
supported in part by the Bundesministerium f\"ur Forschung und Technologie
(BMFT) under contract number 06 HD 710.

\section*{Figure captions}
{\bf Fig. 1:} The inelastic interaction of a quark $q$ and a antiquark
$\overline q$ with $n$ nucleons of target nucleus.

\medskip
\noindent
{\bf Fig. 2:} The ratios $R(A/B)$ of the Drell-Yan cross sections
$d\sigma^{hA}_{DY}/d^2p_T$ and $d\sigma^{hB}_{DY}/d^2p_T$ as a function
of the transverse momentum $p_t$ of the D-Y pair. The data from
experiments with various projectiles, targets and energies are compared
with the calculation based on eq. (\ref{aj}).
\end{document}